# Can in-home laboratories foster learning, self-efficacy, and motivation during the COVID-19 pandemic?-A case study in two engineering programs

Jonathan Álvarez Ariza

Program of Technology in Electronics, Engineering Faculty, Corporación Universitaria Minuto de Dios-UNIMINUTO, 111021 Bogotá, Colombia. E-mail: jalvarez@uniminuto.edu

**Abstract**
The COVID-19 pandemic has represented a challenge for higher education in terms to provide quality education despite the lockdown periods, the transformation of the in-person classes to virtual classes, and the demotivation and anxiety that are experimented by the students. Because the basis of engineering is the experimentation through hands-on activities and learning by doing, the lockdown periods and the temporary suspension of the in-person classes and laboratories have meant a problem for educators that try to teach and motivate the students despite the situation. In this context, this study presents an educational methodology based on Problem-Based Learning (PBL) and in-home laboratories in engineering. The methodology was carried out in two phases during 2020, in the academic programs of Industrial Engineering and Technology in Electronics with ($n=44$) students. The in-home laboratories were sent to the students as part of "kits" with the devices needed in each subject. Besides, due to the difficulties in monitoring the learning process, the students made videos and blogs as a strategy to reinforce their learning and evidence the progress in the courses. The outcomes of the methodology show mainly the following points: (1) An improvement of the academic performance and learning of the students in the courses. (2) A positive influence of the usage of in-home laboratories in motivation, self-efficacy, and reduction of anxiety. (3) Positive correlations between the usage of in-home laboratories, the blogs and videos, and the teacher's feedback for learning, motivation, and self-efficacy. Thus, these results evidence that other alternatives that gather the cognitive and affective learning domains can emerge from engineering to deal with the educational problems produced by the crisis periods. Although the methodology and the lessons learned in this study are applied to the pandemic period, they can be extended in engineering education to the post-pandemic period.

## 1. Introduction

The COVID-19 pandemic has represented a big challenge for higher education institutions around the world that struggle to keep the classes despite the restrictions of mobility to guarantee the safety of students, professors, and administrative staff. This situation has had an important impact on the engineering faculties in several aspects. The first one is the quick transition between in-person classes towards an Emergency Remote Teaching (ERT) format [1]–[5], continuing with the quality requirements in education. Although this transition was agile and it was provided by resources, especially regarding the Information and Communication Technologies (ICTs), there are many questions about the implications that they represent for the educational process of the students, considering that in some cases this transition has been taken as the simple use of online platforms as a way to deliver instructional contents [6]–[9]. In this sense, ERT differs from online learning since it is a temporal method of learning and teaching due to crisis circumstances. As Hodges et al. [1] describe, the primary objective of the ERT is not to recreate a robust educational system but to be a method to support the learning needs of the students under the crisis period. According to UNESCO [5], the socioeconomic factors, anxiety, problems with ICT and internet connectivity, social isolation, and keeping the regular scheduling are impacts produced by the pandemic in the world but with special emphasis in Iberoamerica and North America regions. The second one is related to the mobility restrictions in laboratory practices and experimentation that are substantial components in engineering. Laboratories represent the basis of the experimentation in engineering, and they have formed an active part of the curricula. Students experiment and construct their meanings with hands-on activities, which are the essence of science learning [10]. With the lockdown periods, the laboratories and practices have been reduced in extreme, and they have tended to be replaced by simulators or remote laboratories. However, not all higher education institutions have had important budget resources to cope effectively with the pandemic, and many students could result affected in the teaching continuity [10]. Finally, the third one is concerning the general question for educators about how to promote and guarantee problem-based learning and collaboration between the students given the current scenario under the COVID-19 pandemic. While infrastructure, faculty support, and budget components in universities are important factors, also the educational methodologies and the role of the educators to promote learning, reflection, and participation of the students in the period of crisis.

In this context, this study proposes a Problem-Based Learning (PBL) methodology [11] with in-home laboratories and blogs to support the learning of the students in two programs of engineering, namely, Industrial Engineering and Technology in Electronics in the subjects of automation, industrial instrumentation, and introduction to electronics. The in-home laboratories were sent as "kits" with devices such as multimeter, protoboard, voltage source, Arduino, sensors, among other devices in order to the students developed the proposed tasks in the courses. The kits used several open-hardware elements such as Arduino and 3D robots because of the accessibility in terms of cost and facility of construction. Regarding open-source hardware and software platforms in engineering and their implications for learning and research, see the works of Joshua Pearce [12]–[15].

The methodology was carried out with (*n=44*) students during 2020. The students of industrial engineering are from the eighth semester while students of technology in electronics are distributed in the first and fifth semesters. Students were encouraged to work in groups of between two and three people to foster collaboration. Due to the impossibility of doing in-person classes and the challenge of monitoring the learning of the students, it was selected the work with blogs whose aim was twofold. Firstly, the blogs allow the students to register the advance of their learning in entries that encompassed the topics and concepts in the courses. Blogs can work with different types of resources such as videos, documents, and diagrams which can help the students learn better [16]–[18]. Secondly, for educators, the monitoring of the activities made by the students and the respective feedback is much easier. In the methodology, the students made several videos that had at least two components: (1) an explanation of the concepts addressed in the tasks or activities, and (2) a detailed account by each group member that synthesized *how* through the devices in the in-home laboratories, he/she solved the proposed problems in the courses. Although an accepted and extended practice to support the learning of the students is the usage of simulators in the pandemic, the hands-on activities directly with hardware devices help the students to experiment actively and learn by doing which is an educational cornerstone concept in the areas of Science, Technology, Engineering and Mathematics (STEM). The real interaction with devices that will be used in the future labor life of the students, in many cases, cannot be replaced using only simulators since they were designed to support the educational process under controlled variables and environments. So, the interest with the methodology is that the students experiment actively and learn by doing in other spaces of learning different to the university, e.g., from their homes which can help to support their educational process.

The methodology was divided into two phases. In phase one (pilot study) was analyzed the perceptions of the students regarding the usage of in-home laboratories and blogs with *n=14* students. With the conclusions of this phase, the improvement of the instruments, and the inclusion of complementary variables that influence motivation such as self-efficacy or anxiety, phase 2 was performed with *n=30* students. The study adopted the research approach of embedded design [19], [20], where the questions in the surveys had both closed-ended and open-ended questions that were examined in parallel to understand the perceptions of the students. To analyze the collected data, it was considered the academic performance with the students' grades at the beginning and the end of the semesters; the results of the surveys in both phases in Likert Scale; and an evaluation rubric [21] for the videos and blogs made by the students whose structure was checked by an education expert. The rubric was divided into the categories of organization, content coherence, concept comprehension, concept application, and teamwork. Data were analyzed using descriptive statistics, Cronbach's alpha for the surveys' reliability [22], Wilcoxon signed-rank, and Mann-Whitney U tests [19] to find statistical significances between the constructs of the study in the courses, and Pearson's correlation matrix to find relationships in the assumed categories. The analysis was performed employed the software IBM SPSS statistics v.23. The outcomes show the following points: (1) an improvement in both learning and the academic performance of the students through the methodology. (2) A positive correlation between learning with in-home laboratories, motivation, self-efficacy, reduction of anxiety, and teacher's feedback. (3) Reduction of nervousness in the delivery of assignments and exams. (4) Reinforcing of the learning through the videos and blogs. (5) Finally, a good acceptance and positive comments by the students about the methodology and its implications for their learning.

The remainder of the paper is organized as follows. Section 2 exposes the background of this study. Section 3 describes the educational context of the methodology. Section 4 depicts the research questions, participants, and the instruments adopted in this study. Section 5 shows the results of the study which are addressed according to the posed research questions. Section 6 illustrates the lessons learned and the implications of them for engineering education. Section 7 depicts the limitations and further work of this research. Finally, section 8 outlines the conclusions for this study.

**2. Background**

This section discusses the theoretical underpinnings that shape this study that are mainly focused on Problem-Based Learning (PBL) and self-efficacy.

*2.1 Problem-Based Learning (PBL)*

PBL is an educational approach in which learning is organized around the resolution of meaningful problems that demand investigation and explanation from the students [11], [23], [24]. The work of the students under this approach is usually in small groups with the accompaniment of a tutor or teacher which is a facilitator of the educational process [25]. Among the features of PBL are the following [11], [23]:

1. Usage of problems as starting point to learning. Problems could be discipline-specific or multidisciplinary. Also, the problem must be authentic, real-world, and ill-structured.
2. The students work collaboratively the main part of the time usually in small groups. It is required that the students can learn in a self-directed way, identifying their knowledge gaps.
3. Flexible guide of the tutor or instructor. The tutor presents the problem situation.
4. The number of lectures is limited.
5. The assessment could be knowledge-based or process-based.
6. Learners or students have access to all information available that is involved in the problem analysis.

Concerning the definition of the notion of problem inside PBL, Jonassen [26] has indicated that this is an unknown entity in a context, e.g., social, technical, or cultural. So, two classifications of problems have emerged: *well-structured problems and ill-structured problems*. The well-structured problems require the application of a limited number of concepts, operations, instructions, which are organized and sequenced, e.g., by the students. As for the ill-problems, they are formulated with a non-conventional situation where the methods are not known and there are multiple solutions to address them. Moreover, with the ill-structured problems is expected that the students recognize their current knowledge gaps, even though, to solve the problem prior knowledge is not required [23]. In general terms, PBL involves five phases [23], [27]: (1) Problem presentation, and exploration. (2) Problem analysis and identification of knowledge gaps. (3) Information gathering. (4) Discussing ideas in group based upon the acquired knowledge. And (5) Reflection of the process based on the tutor or teacher's feedback.

Tutor or facilitator for the students acts as scaffold [23], [25]. That is, facilitators promote the learning of the students, but this function is progressive relying on the advantages that they could have. Similarly, two characteristics are expected in the facilitators: subject matter experience, and cognitive and social congruence [23]. In the first one, the facilitators must have experience in the subject that they teach to help to clarify the different doubts of the students in the process. Besides, when facilitators possess experience, they can guide the educational process more adequately, creating productive learning environments. In the second one, facilitators should have interpersonal qualities that bring them closer to students. This closeness motivates the students to learn and clarify their doubts. In synthesis, these are the main features of the PBL methodology that have been considered in the structure of this study.

*2.2 Self-efficacy*

Self-efficacy or perceived self-efficacy refers to the personal beliefs about possessing capabilities to cope with determined situations and get the desired results in these [28]–[30]. Then, self-efficacy is not associated with the number of capabilities that one could have, but with the beliefs about these. Given that self-efficacy is compounded by a set of beliefs, these could come from four sources [28], [31]:

- Enactive mastery experiences: In these, the persons reinforcing their perceptions with the overcoming of situations that normally are generated in everyday life. According to the interpretation of the success or the failures, the person could have a better or worst performance in the activities. Thus, attentional,

- physical, and emotional factors can influence the variation in the quality performance [28]. Also, the resilience to deal with the problems helps to strengthen the perceived self-efficacy.
- Vicarious experience: This arises from the comparisons of the own performance with others that typically are similar. e.g., peers in a university course, coworkers, etc. Thus, these social comparisons acting a primary factor in the self-appraisal of capabilities.
- Verbal persuasion: Social persuasion serves as a mean to shape the beliefs about the own capabilities. Due to that in many activities, the persons by themselves cannot evaluate their capabilities, the verbal persuasion functions as a source to strengthen the beliefs. Similarly, in verbal persuasion, the confidence in the performance is mediated through the perceived credibility of expertness or persuaders, for instance, in the case of a tutor or teacher. As Bandura [28] observes, some skills require intricate knowledge about the development of proficiency in given pursuits, which can be facilitated by an expert or persuader.
- Physiological and affective states: Persons read their physiological situations as signs of vulnerability or dysfunction. Anxiety, desperation, demotivation are states that influence self-efficacy and performance. So, complex tasks go accompanied by high emotional activation and the arousal produced by them can serve as a facilitator or a debilitator for the performance.

Although the previous constructs arise from psychology, these have important relevance in the educational process of the students even more due to the pandemic situation. For instance, the affective states of anxiety and demotivation that students have assumed could impact learning, self-efficacy, and academic performance. Therefore, it is needed the creation of educational methodologies that not only are focused on the cognitive learning domain [32]–[34] but also include the affective learning domain [35], for instance, the ones mentioned in [36]–[38]. In this way, a complementary aim with this work is to explore how self-efficacy, anxiety, or motivation are involved in the learning of the students.

**3. Educational context**

The methodology was conducted in two phases at the University Corporación Universitaria Minuto de Dios-UNIMINUTO in the engineering faculty during the first and second semesters of 2020. The first phase matches with the starting of the pandemic and the lockdown period decreed by the Colombian authorities. To preserve the health of the students, professors, and administrative staff, the university canceled all in-person classes and laboratories in March of 2020 and migrated to the virtual learning modality. Thus, in this period the students experienced the transition from in-person classes to virtual classes utilizing Google Meet, Microsoft Teams, Zoom, Moodle, and electronic simulators, e.g., Proteus VSM or EDA playground. Because the programs of engineering and technology require hands-on activities that support the basis of the curricula, it was decided to construct an educational methodology with PBL to deal with the educational challenges produced by the pandemic. For that, several kits were designed and sent to the home of the students with low-cost hardware devices needed in each subject. With these kits, the students made the in-home laboratories. In parallel, it was started a pilot study to analyze the implications of the methodology and its acceptance by the students (*n=14*) in the subject of introduction to electronics that encompasses topics in basic circuits, handling of laboratory instruments (power supply, multimeter, oscilloscope), introduction to programming with hardware elements, and fundamentals of Printed Circuit Board (PCB) design. This subject is offered to first-semester students. Because the ERT format requires a transformation of the classroom spaces, the methodology was developed considering the following aspects:

1. The students should work in groups of two or three people.
2. Each assignment, project, or exam must count with an explanation video. Each student of the group must do a video that evidences the learned concepts in the subject with the help of the materials in the kit.
3. The videos and the development of the different tasks in the courses must be uploaded to a blog created in any platform such as WordPress, Blogger, among others.
4. The classes were transformed into spaces to debate ideas, doubts, and problems with the tasks in a virtual modality.
5. To clarify doubts about the proposed activities and problems, some in-person laboratories were scheduled during the semesters according to the guidelines of the university and the national ministry of education.

The aim to use videos for the students was twofold. On one hand, with the videos, the students can identify if the concepts were adequately understood because they made several descriptions about how to solve the posed problems with the help of the devices in the kit. On the other hand, the videos or blogs can help instructors or professors to monitor the learning process of the students which is difficult under the modality of virtual classes. The second phase was performed in the second semester of 2020 with the conclusions, insights, and improvement of the instruments of the pilot study. These improvements were more meaningful and successful for the learning process of the students which was demonstrated by the educational outcomes derived from the second phase. In the second stage participated students of the career of Industrial Engineering and Technology in Electronics in the subjects of automation ($n=26$) and industrial instrumentation ($n=4$). Regarding the first subject, it encompasses topics in basic circuits, programming, and fundamentals of industrial devices. The second subject addresses topics in the fundamentals of digital and analog sensors for industrial applications. With the same rationale of phase one, the kits were sent to the students with the materials needed in each subject. Because most of the students have a job, they took their classes in the night schedule from 6:00 pm to 10:00 pm, in the subjects of introduction to electronics and automation ($n=40$). Academically, each subject is divided into two quarters and a final exam with a duration of 16 weeks. In both phases, the in-home laboratories started to be employed from the second quarter to compare the methodology with the typical classes in a virtual modality in the first quarter. To prevent differences in the courses due to uncontrolled variables, for instance, about the different instructional methods, the same teacher imparted the classes in all the courses. The complementary information about the methodology and the hardware and software components employed can be found in the appendix section. Considering these aspects, the next section discusses the research questions and the followed method in the study.

## 4. Method
### 4.1 Research Questions

For the study, several Research Questions (RQ) were proposed concerning the usage of the in-home laboratories, the videos and blogs, and their impact on learning, motivation, self-efficacy, and anxiety. The list of these RQs is illustrated as follows:

**RQ1.** Does the usage of in-home laboratories impact positively in learning and academic performance?
**RQ2.** Does the usage of in-home laboratories influence motivation, self-efficacy, and anxiety?
**RQ3.** Can the construction of videos and blogs influence positively on learning and teamwork?
**RQ4.** What are the relationships between learning with in-home laboratories, self-efficacy, motivation, anxiety, and instructor's feedback?

Searching to answer these RQs, the research approach selected for the study was an embedded design [19], [20] where the surveys provided to the students had a gather between quantitative and qualitative questions to complement the results. Therefore, the report on the results contains a dialog between quantitative data and the perceptions of the students since these can indicate if the methodology was adequate and relevant.

### 4.2 Participants

In phase one, $n=14$ students participated in the methodology with an age average of 21 years old. For the second phase, $n=30$ students participated with an age average of 24 years old. In both phases, the gender distribution was 74.5% male and 25.5% female students.

### 4.3 Instruments

*Phase one (pilot study)*: To collect the perceptions of the students, a survey was administered with 9 closed-ended questions on a Likert scale in the range (1) Strongly disagree; (2) Disagree; (3) Agree; (4) Strongly agree, and two open-ended questions. Likewise, with the open-ended questions, it was collected both the positive aspects and the elements to improve with the methodology. Regarding the closed-ended questions, these were divided into three categories with the following Cronbach's alpha ($\alpha$) values:

- Learning and Teamwork with Videos and Blogs (LTVB) (5 items, Q1-Q5, $\alpha=0.703$).
- Learning with in-Home Laboratories (LHL) (3 items, Q6-Q8, $\alpha=0.85$).
- Motivation with in-Home Laboratories (MHL) (1 item, Q9).

The survey was answered by all students (*n=14*, 100%) and it aimed to explore the acceptance of the methodology concerning the in-home laboratories and the development of the videos and blogs.

*Phase two*: The instrument in phase one was improved to collect more information about the implications of the usage of the in-home laboratories, blogs, and videos in the educational process. Then, a survey with 13 closed-ended questions on a Likert Scale in the range (1) Strongly disagree; (2) Disagree; (3) Agree; (4) Strongly agree, and 2 open-ended questions was applied to the students. For the survey, four categories were designed:

- Learning with in-Home Laboratories (LHL) (5 items, Q1-Q5, α=0.833).
- Learning and Teamwork with Videos and Blogs (LTVB) (3 items, Q6-Q8, α=0.869).
- Motivation, reduction of Anxiety, and Self-efficacy (MAS) (4 items, Q9-Q12, α=0.772).
- Instructor's Feedback of learning process (IF) (1 item, Q13).

Although the Cronbach's alpha (α) values in the survey of the pilot study indicate good reliability of the instrument, several answers provided by the students, in special in the category of MHL, indicated that other variables can affect motivation in the courses. Thus, for the instrument in phase 2 was added the constructs of self-efficacy and anxiety in the category of MAS. By the same token, this study explored if the instructor's feedback influenced other categories of the survey. These constructs take more relevance given the current context of the ERT by the COVID-19 pandemic. Aspects such as the limitations in the social interactions, the lockdown periods, and the modification of the in-person classes by virtual classes could affect the motivation, learning, and engagement in the courses.

Similarly, to phase one, the open-ended questions collected the information of positive perceptions and aspects to improve in the methodology. In this case (*n=28*, 93.33%) of the students answered the survey. Moreover, due to the number of videos and blogs made by the students, an evaluation rubric for these was constructed with the categories of organization, content coherence, concept comprehension, concept application, and teamwork. Each blog and video made by the students was analyzed under this rubric whose structure was assessed by an expert in education. The link for the rubric and the surveys in each phase can be found online in the appendix section.

### 4.4 Analysis

Data were collected by the same teacher/researcher in the courses which facilitated the application of the instruments. Firstly, the value of Cronbach's alpha (α) was calculated to examine the reliability of each instrument. Also, it was performed descriptive statistics in terms of mean (*M*) and Standard Deviation (*SD*) in the categories of the surveys described above. For the rubric was computed the average of the score obtained for the students in each one of the mentioned categories. Secondly, to find significant differences in the academic grades of the students with the methodology during the quarters and final exam in the courses, it was carried out an inferential statistical analysis, employing the Wilcoxon signed-rank test. Thirdly, to analyze statistical differences in motivation between the courses in the first and second semesters of 2020 was performed a Mann-Whitney U test. Fourthly, a Pearson's correlation analysis was generated to discover relationships in the categories of the survey of phase two, e.g., between the usage of the in-home laboratories and the motivation or anxiety. Finally, the comments of the students with the open-ended questions were studied to contrast the quantitative results in each phase of the study. The results and discussion will be presented according to the proposed Research Questions (RQ).

## 5. Results and discussion

### 5.1 *RQ1*: Does the usage of in-home laboratories impact positively in learning and academic performance?

The descriptive statistics for the category Learning In-Home Laboratories (LHL) in the surveys of the pilot study (*M=3.64, SD=0.48*) and stage 2 (*M=3.82, SD=0.3*) suggest that students had a good perception of the influence of the in-home laboratories in their learning. As for stage 2, the Mean and SD show an improvement of the results in comparison with the pilot study. This fact is mainly because the students in phase 2 counted with a major number of educational resources, e.g., videos, tutorials, and lectures that accompanied their experiences with the in-home laboratories as well as the tutoring process was stronger than in the pilot study.

To contrast these results with the academic performance of the students, it was generated a Wilcoxon signed-rank test with their grades in the First Quarter (FQ) and Final Exam (FE) to found if learning with the methodology and the in-home laboratories was statistically significant. The rank of the students' grades was from 1 to 5. As described, in the FQ the classes were entirely virtual without any in-home laboratory, only assignments, and simulations in the courses. It was selected the non-parametric Wilcoxon signed-rank test because data in some courses do not fit with a normal distribution according to the Kolmogorov-Smirnov test whose results for the subjects of Introduction to Electronics, and Automation in the FQ and FE were the following:

- Introduction to Electronics (FQ): $D(14) = 0.21$, $p=0.082$.
- Introduction to Electronics (FE): $D(14) = 0.15$, $p=0.2$.
- Automation (FQ): $D(26) = 0.17$, $p=0.049$.
- Automation (FE): $D(26) = 0.24$, $p=0.001$.

In this case, the students' grades of Automation do not meet the normal distribution criteria since $p<0.05$. Regarding industrial instrumentation due to the small number of students ($n=4$), the Kolmogorov-Smirnov test is not suitable. Thus, it was performed the Wilcoxon signed-rank test whose results show for the course of Automation, a statistically significant difference between the FQ (*mean rank*=11.29, *sum of ranks*=79) and FE (*mean rank*=14.32, *sum of ranks*=272), $T=79$, $z=-2.452$ (corrected for ties), $N$-Ties=26, $p=0.014$. Similarly, for the subject of Introduction to Electronics, FQ (*mean rank*=3, *sum of ranks*=9) and FE (*mean rank*=8.2, *sum of ranks*=82), $T=9$, $z=-2.551$ (corrected for ties), $N$-Ties=14, $p=0.011$. In both courses the number of ties was zero, indicating an improvement of the grades in all students between the FQ and FE. As for the course of industrial instrumentation, in the FQ ($M=4.05$, $SD=0.64$) and in the FE ($M=4.25$, $SD=0.5$), which demonstrates an improvement in the academic performance. In addition, to compare the students' grades distribution in the courses in the FQ and FE, Fig.1 illustrates a boxplot with this information.

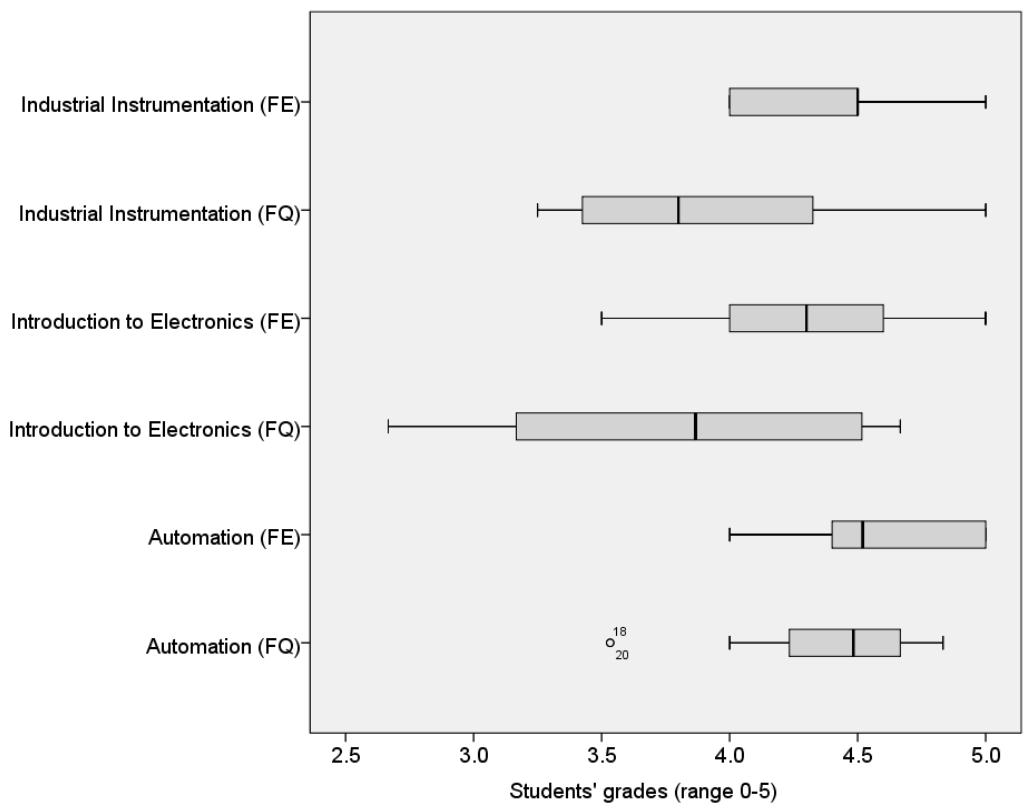

**Fig.1 Comparison of the students' grades in the courses between the First Quarter (FQ) and Final Exam (FE). *N*=44**.

Concerning Fig.1, a better improvement was achieved in the subject of introduction to electronics that is offered to students of the first semester. In this subject, it was used a mobile application with the in-home laboratories to the students learned to program and interact with hardware devices and software, and this fact was more relevant for the students. Pertaining to the previous aspects, some students commented the following:

*S1. The usage of kits was an excellent pedagogical tool that motivated me and allowed me to have experience in the professional formation, reducing the limitations with the laboratories due to the pandemic.*
*S2. The class was very didactic thanks to the kits since they allowed me to experiment and gain experience with the topics of the course.*
*S3. I consider that was interesting the usage of kits to address the different topics in the course.*
*S4. It was very interesting to have a mobile application to be able to program outside the home or the laboratory.*
*S5. The mobile platform that we employed with the in-home laboratories was very useful to understand programming in C language.*

In phase 2, students were asked to indicate the most relevant aspects for learning in the courses with the methodology. Fig.2 shows the distribution of these answers. Students indicated the experimentation with in-home laboratories (54%) and the reinforcing of the learned concepts through the blogs and videos (21%) as the most relevant aspects of the methodology. In sum, these aspects suggest that the in-home laboratories and the methodology in the courses had an impact on learning and academic performance in the students, corroborating the assumption in RQ1.

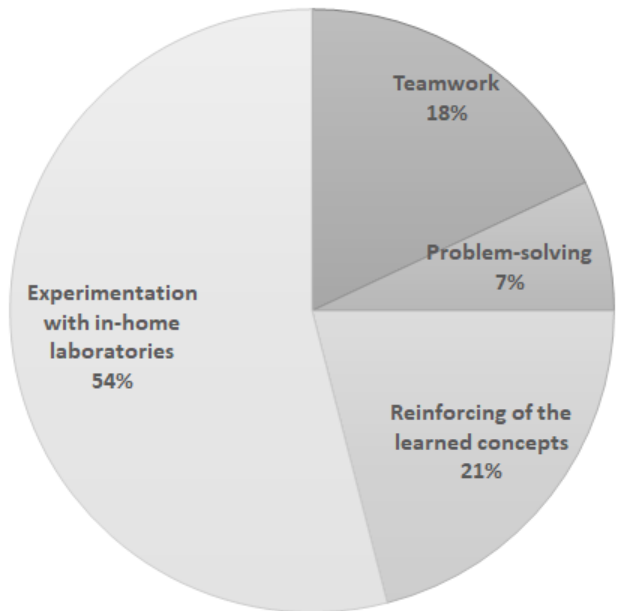

**Fig.2 The most relevant aspects indicated by the students in phase 2 of the methodology. *N*=28.**

**5.2 RQ2.** *Does the usage of in-home laboratories influence motivation, self-efficacy, and anxiety?*

As for the pilot study, the category Motivation with in-Home Laboratories (MHL) was evaluated by the students in the survey with ($M$=3.21, $SD$=0.89). This result is because some students indicated that the methodology did not improve their motivation in the course. Factors as the quick transition from in-person classes to virtual classes, the tutoring process, and the limitation of the laboratory practices could influence this result. So, in phase 2 was inquired about other variables that could be related to motivation, e.g., self-efficacy or anxiety. As described, self-efficacy refers to the beliefs that one could have concerning the capabilities to get the desired result in a particular field, and it has a close relationship with motivation. Similarly, anxiety plays a role as a physiological state that could influence the developed beliefs about self-efficacy. Then, Q9 inquired for motivation, Q10-Q11 for self-efficacy, and Q12 for reduction of anxiety through the methodology. These questions belong to the category (MAS) in the survey. For motivation ($M$=3.75, $SD$=0.44), self-efficacy ($M$=3.64, $SD$=0.42), and reduction of anxiety ($M$=3.6, $SD$=0.49). All students in phase 2 indicated that the methodology helped to improve their motivation in the courses and allowed the classes were more didactic and interesting. Furthermore, students stated that in comparison with the start of the courses, they felt more confident concerning the developed skills for problem-solving. Likewise, the students indicated that the methodology made easier to understand the addressed topics and reduced the anxiety in the proposed exams and activities. To compare the outcomes of the motivation in the first and second semester of 2020, a Mann-Whitney U test was conducted on the data ($n=42$). The results show statistical difference between groups since $p<0.05$, (2020-I: *mean rank*=16.5,$n=14$; 2020-II: *mean rank*=24, $n=28$), $U=126$, $z=-2.224$, $p=0.026$. These differences in motivation are due to the factors that were described between the pilot study and phase 2. These descriptive and inferential outcomes suggest that the in-home laboratories and the methodology influenced the motivation, self-efficacy, and reduction of anxiety in the students.

### 5.3 RQ3. *Can the construction of videos and blogs influence positively on learning and teamwork?*

As mentioned, the interest in blogs and videos was twofold. On one hand, these aimed that the students reinforced the concepts learned in class since they had to explain and describe *how* to solve diverse problems with the kits. This methodology created a cycle in which the students had to review and analyze if the concepts were adequately learned. It is worth mentioning that this process depends on the feedback provided by the teacher to the students as we will see in the description of the relationships between constructs in RQ4. Moreover, the students could edit their blogs collaboratively from their homes or jobs. For the pilot study and phase 2, the students evaluated the category Learning and Teamwork with Blogs and Videos (LTVB) with ($M=3.58$, $SD=0.36$) and ($M=3.63$, $SD=0.39$), respectively. The students made several comments as follows:

*S1. With the blog was evidenced the appropriation of the topics in the subject, which help to progress and understand some concepts clearly.*
*S2. The blog allowed me to share information with my peers and reinforce the handle of ICT.*
*S3. With the blog, I was able to investigate and reinforce the concepts learned in class.*
*S4. I think that some positive aspects of the blog were the investigation of complementary concepts, work collaboratively with my peers, and practice with ICT.*
*S5. I think that at the starting of the methodology was difficult to adapt me to the tool Google Sites for the construction of the blog.*

The students indicated in both phases that the blogs and videos allowed them to work collaboratively despite the lockdown periods. In addition, some students manifested that the tools to make their blogs were difficult to handle at the start of the methodology. Concerning Fig.2, (21%) of the students stated that the blogs and videos helped to reinforce the topics in the courses. Table 1 illustrates some examples of blogs developed by the students (in Spanish) with their respective URLs.

**Table 1**. Examples of blogs constructed by the students.

| Blog example | Course | Access link |
|---|---|---|
| 1 | Automation | https://bit.ly/2NJbHh6 |
| 2 | Automation | https://bit.ly/3dyZcPW |
| 3 | Automation | https://bit.ly/3dviaXS |
| 4 | Introduction to electronics | https://bit.ly/3sqzAt5 |

On the other hand, the videos and blogs facilitated the monitoring of the learning process of the students. Because the classes gathered hands-on activities, a manner to know if the students learned and applied the concepts with the in-home laboratories was through videos and blogs. In phase 2, 10 blogs were made by the students, and they were evaluated by the designed rubric in its categories in the range 1 to 5. As mentioned, the rubric evaluated the organization, content coherence, concept comprehension, concept application, and teamwork. The overall scores for the blogs were ($M=4.58$, $SD=0.33$), which is an indicator that students made and corrected properly their blogs and videos with the help of the teacher's feedback. The described results show that videos and blogs can contribute to enhance the learning of the students, but it depends on the monitoring and feedback that the teacher does to the students.

**5.4 *RQ4*.** *What are the relationships between learning with in-home laboratories, self-efficacy, motivation, anxiety, and instructor's feedback?*

To find the relationships between the categories of the survey and the teacher's feedback in phase 2, it was calculated a Pearson's correlation matrix as it is illustrated in Table 2. In the table, the most significant positive correlations between categories (over 0.7, *p*<0.01) are highlighted in bold. As described, the survey was divided into the categories of Learning with in-Home Laboratories (LHL), Learning and teamwork with Videos and Blogs (LTVB), Motivation, Anxiety, Self-efficacy (MAS), and the teacher's or Instructor's Feedback (IF) in the learning process. To find the scores in each mentioned category were averaged the values of the students in those before applying Pearson's correlation.

Table 2. Pearson's correlation matrix for the categories in the survey of phase 2. *N*=28.

| Categories | Motivation | Self-efficacy | Anxiety reduction | LTVB | LHL | IF |
|---|---|---|---|---|---|---|
| Motivation | 1 | | | | | |
| Self-efficacy | **0.73**** | 1 | | | | |
| Anxiety reduction | **0.718**** | **0.786**** | 1 | | | |
| LTVB | 0.667** | **0.701**** | 0.564** | 1 | | |
| LHL | **0.77**** | **0.798**** | 0.605** | 0.579** | 1 | |
| IF | **0.808**** | **0.737**** | 0.58** | 0.445* | **0.794**** | 1 |

\** Correlation is significant at the level of *p*=0.01. * Correlation is significant at the level of *p*=0.05.

According to Bandura [28], [30], perceived self-efficacy and its beliefs are associated with the enactive mastery experiences, vicarious experiences, physiological and affective states, and verbal persuasion. For the data in the courses, the anxiety reduction has a positive relationship with motivation ($r(28)=0.718$, $p<0.01$) and self-efficacy ($r(28)=0.786$, $p<0.01$), which is agreed with the postulates indicated by Bandura regarding the physiological and affective states. Similarly, the category LHL has positive correlation between motivation ($r(28)=0.77$, $p<0.01$) and self-efficacy ($r(28)=0.798$, $p<0.01$). So, the students indicated in the survey that the proposed problems with the in-home laboratories benefited their skills in comparison with the start of the course concerning dealing with problem-solving. By the same token, the improvement of the motivation reduced the anxiety with the exams, laboratories, and activities posed in the courses.

One interesting point in the correlations was the influence of the teacher's or instructor's feedback on motivation, self-efficacy, and LHL. Since one of the sources for self-efficacy is the verbal persuasion guided by the credibility and expertness of persuaders, the teacher's feedback provided to the students demonstrated a positive correlation with self-efficacy ($r(28)=0.737$, $p<0.01$), motivation ($r(28)=0.808$, $p<0.01$) and LHL ($r(28)=0.794$, $p<0.01$). Also, the role of the teacher as facilitator of the educational process according to PBL [25] is supported by the previous results. At last, LTVB has a positive relationship with self-efficacy ($r(28)=0.701$, $p<0.01$). This fact could be influenced by the perception of the students about the blogs and videos that reinforced their learning in the courses. As it is depicted in Fig.2, (21%) of the students recognized as most relevant this feature of the methodology. Previous correlations show that the results depicted in the previous RQs are reliable, coherent, and respond to the purposes of this study.

**6. Lessons learned and implications for engineering education**

While diverse overall policies have arisen from the macrolevels (governments, ministries, industries, universities, etc.) among other institutions from the economic and technical resources, it seems to be that a low number of educational options have been proposed by educators to deal with the educational effects produced by the pandemic in the engineering field. Perhaps, this is one of the fields more impacted in education due to factors such as the limitations in the laboratory practices and hands-on activities, the problems with the access to the internet experimented by the students, demotivation, anxiety, among other factors. So, the current situation under the COVID-19 pandemic requires a transformation of the classroom spaces to offer high-quality education to the students. Although the previous aspects in the RQs were analyzed under the lens of descriptive and inferential statistics and they suggest positive outcomes, these have been achieved through a careful planning and design of an educational methodology. Taking into account the results for the methodology, the following lessons learned were concluded:

1. *Classroom Transformation*: Virtual sessions were transformed in active spaces to debate ideas and solve doubts concerning the topics and in-home laboratories. Several sessions were focused on the theoretical aspects required by the students. Also, several tutoring sessions were developed with the students. This

      proposal utilized PBL since this methodology is more adequate to work with the students through the in-home laboratories, blogs, and videos.

2. *Employment of in-home laboratories:* The in-home laboratories were thought to support the experimentation and learning by doing of the students, and to reduce the health risk of the pandemic. Some classes are more suitable to use in-home laboratories than others, in special, those that do not require expensive laboratory equipment. However, teachers should analyze what classes could use these laboratories and what hardware devices and software are required for them.
3. *Use of simulators to support laboratories*: Simulators can support the learning process, but the simulators cannot convert into unique resources in classrooms. Other alternatives according to the selected educational methodology should be explored, e.g., in-home laboratories, mobile learning platforms, remote laboratories, among others.
4. *Learning or emotions*: Sometimes in engineering, there is a dichotomy between learning or emotions. The current situation with the pandemic needs other points of view that complement the cognitive learning domain. In this study, for instance, motivation, reduction of anxiety, and self-efficacy are correlated with learning and academic performance. Educators should consider *these other* affective variables that influence the educational process of the students.
5. *Limit excessive workload*: Since the students are in their homes, this does not implicate that they can solve many assignments, laboratories, or exams. It is needed to plan carefully the tasks that will be solved by them to prevent demotivation, excessive cognitive load [39], or breaches that could affect the engagement in class.
6. *Teacher's Feedback*: A fundamental element in the methodology is the teacher's feedback. Because students are working with in-home laboratories, they need to know if the learning of the concepts, their application, and the reporting process with the videos and blogs are correct. With the feedback, the students can identify their gaps and errors with the possibility to correct them, which impacts the perceived self-efficacy.
7. *Blogs and videos*: In the proposal, the videos and blogs were used to reinforce learning in the students, but also as a method to monitoring their advancement in the courses. The initiative arose from an overall question in the current context and even in the post-pandemic period with the limitation of the in-person classes: *How does the teacher or instructor know or evidence that the students are learning?* If the practice and learning by doing in engineering are essential components of the curricula solely the written exams, assignments, or simulators could not be enough.

Although these lessons have been thought for the pandemic, they can be taken in consideration for the post-pandemic period. The COVID-19 pandemic has opened new questions about how engineering education has been assumed, and it allowed to rethink the role of the educators and the transformation of the educational settings. In sum, these are some lessons learned with the methodology and the employment of the in-home laboratories that can help educators that are interested in creating educational methodologies that support learning.

## 7. Limitations of the study and further work

Although in the subject of industrial instrumentation, it was considered a sample of *n=4* students, which is low, the methodology was tested in all courses, allowing to enhance the learning process since this fact was a primary goal in the study. Therefore, even though the sample was small in this course, the students were benefited from the methodology with its educational implications. In addition, one purpose of this study was to explore the constructs of self-efficacy, reduction of anxiety, and teacher's feedback, and their influence on learning, motivation, and academic performance. In this exploratory sense, the number of questions in the survey regarding these aspects in phase 2 is reduced, searching to expand their number in future studies that require it. Even so, the proposal and its impacts were analyzed during 2020 in two phases with a pilot study, which helped to improve the instruments and the variables to consider. Finally, it was deemed three subjects in the engineering curricula with the methodology to analyze the implications of it in the educational process. With the returning of a normal situation, the in-home laboratories will be maintained to take advantage of the additional spaces different to the classrooms in which the students can learn. So, in the methodology, the home spaces were feasibly employed to learn and experiment which expand the traditional concept of the classroom inside the university. Then, the in-home laboratories will be used to support the learning and experimentation of the students with the progressive returning to normality. Further research will be focused on proving the methodology in several additional courses in the engineering curricula to extend the impact of the presented results.

## 8. Conclusions

This study explored the usage of in-home laboratories with the accompaniment of blogs and videos as a strategy to deal with the effects produced by the COVID-19 pandemic in learning, motivation, self-efficacy, and the reduction of anxiety in engineering. Thus, the different descriptive and inferential statistical analyses and the perceptions of the students indicate that the methodology had a good reception in them, and it contributed to enhance their learning and motivation despite the lockdown periods, or the restrictions in the laboratories that limited the experimentation in the courses. Besides, the usage of blogs and videos reinforced the concepts learned as well as they serve as a method to monitoring the educational process of the students which is difficult due to the virtual classes. Likewise, several variables that affect the academic performance and learning in the students such as motivation, self-efficacy, reduction of anxiety, and teacher's feedback were explored with their relationships and implications. In all courses, the academic performance, motivation, and self-efficacy were improved in the students. While these aspects suggest positive learning outcomes in the cognitive and affective learning domains, these have arisen according to the careful design and planning of a methodology that had the purpose to benefit the educational process of the students despite the limitations of the crisis period. Moreover, the methodology and the lessons learned in the study can be extrapolated to the post-pandemic period, since they were derived from a critical reflection about the role of the experimentation, learning by doing and the teacher in the learning process of the students, and the implications of those for engineering education.

**Appendix section**

Supplementary materials (surveys and rubric) for the phases in the study and the supporting documentation that describes the methodology and the materials (hardware and software) employed for the in-home laboratories with their costs can be found at the Zenodo Repository: https://doi.org/10.5281/zenodo.5512366

**Jonathan Álvarez Ariza** is currently an engineering professor at the Corporación Universitaria Minuto de Dios-UNIMINUTO, Bogotá, Colombia. His research interests are engineering education, control systems, embedded systems, and technology-enhanced learning. He received his B.Eng. in electronics engineering from Universidad Central of Colombia in 2013. Also, he received a technology degree in electronics in 2008 from Corporación Universitaria Minuto de Dios-UNIMINUTO, and a M.Ed. from the Colombian institution Universidad de La Salle in 2018. Since 2008 is a professor in the programs of Technology in Electronics and Industrial Engineering at the Corporación Universitaria Minuto de Dios-UNIMINUTO. He has imparted courses in automation, control systems, circuits, and embedded systems. Besides, he has participated in several research projects in engineering education, and he has published several peer-reviewed articles in conferences and journals.